\documentclass{mystyle}

\def\cpd{\rm{kg^{-1}keV^{-1}day^{-1}}}
\def\mwimp{\rm{m_{\chi}}}
\def\csnospin{\rm{\sigma_{\chi N}^{SI}}}

\def\munu{\mu_{\nu}}
\def\munuebar{\rm{\mu _{\nuebar}}}
\def\nuebar{\rm{\bar{\nu_e}}}
\def\s2tw{\rm{ sin ^2 \theta _W }}
\def\fut{f ( \vec{\rm u} , t )}

\begin{document}

\markboth{Henry T. Wong}
{Ultra-Low-Energy Germanium Detector} 


\title{
Ultra-Low-Energy Germanium Detector for Neutrino-Nucleus Coherent
Scattering and Dark Matter Searches
}

\author{\footnotesize HENRY T. WONG
\footnote{Contact E-mail: htwong@phys.sinica.edu.tw}
}

\address{Institute of Physics, Academia Sinica,
Taipei 11529, Taiwan.}

\maketitle


\begin{abstract}
The status and plans of
a research program
on the development of 
ultra-low-energy germanium detectors with sub-keV
sensitivities are reported.
We survey the scientific goals which include the
observation of neutrino-nucleus coherent scattering, 
the studies of neutrino magnetic moments, as well as
the searches of WIMP dark matter.
In particular, a threshold of 100-200~eV and
a sub-keV background 
comparable to underground experiments were
achieved with prototype detectors.
New limits were set for WIMPs
with mass between 3$-$6~GeV.
The prospects of the realization of full-scale
experiments are discussed.

\keywords{
Neutrino Properties,
Dark matter,
Radiation Detector
}
\end{abstract}

\ccode{PACS Nos.: 
14.60.Lm,
95.35.+d, 
29.40.-n.
}

\section{Introduction}	

A research program in low energy neutrino
and astroparticle physics is being pursued
at the Kuo-Sheng(KS) Reactor Laboratory\cite{texonoprogram}.
The KS laboratory is located 28~m from a 2.9~GW
reactor core and has an overburden of about 30~meter-water-equivalent.
Its facilities were described in Ref.~\cite{texonomagmom},
where limits on neutrino magnetic moments with
a 1.06~kg germanium detector (HPGe) at a hardware threshold of
5~keV were reported. 
The experimental procedures were well-established
and the background were measured.
In particular, a background level of
${\rm \sim 1 ~ event ~ \cpd  }$(cpd) at 20~keV,
comparable to those of underground CDM experiments,
was achieved.
The HPGe data were also used
in the studies of reactor electron neutrinos\cite{rnue}
and for reactor axions searches\cite{raxion}.
Data taking and analysis are being conducted
on a 200~kg CsI(Tl) scintillating crystal
detector array\cite{texonocsi}, 
towards the measurement of neutrino-electron 
scattering cross-section and therefore
the electro-weak angle $\s2tw$.
The future scientific goals are to
develop advanced detectors with
kg-size target mass, 100~eV-range threshold
and low-background specifications
for WIMP dark matter searches\cite{texonowimp} as well as
the studies of neutrino-nucleus
coherent scattering\cite{texonocohsca}
and neutrino magnetic moments\cite{munureview}.

\section{Physics Motivations and Goals}

Results from recent neutrino experiments
provide strong evidence for neutrino oscillations
due to finite neutrino masses and
mixings\cite{pdg06}.
Their physical origin and experimental consequences
are not fully understood.
Experimental studies on the neutrino properties
and interactions can shed light on these
fundamental questions and constrain theoretical models,
from which
unexpected surprises may arise.
It is therefore highly motivated to look for and establish
alternatives of neutrino
sources and detection channels, especially
in regions of parameter space which are
experimentally unexplored.

One of the frontiers is to open the
detector window in the previously unexplored
low energy ``sub-keV'' regime. If experimentally realized,
several important subjects in neutrino and
astroparticle physics can be pursued. They
are discussed in the following sub-sections.

\subsection{Neutrino-Nucleus Coherent Scattering}

Neutrino coherent scattering with the nucleus\cite{cohsca}
\begin{equation}
\rm{
\nu ~ + ~ N ~ \rightarrow ~
\nu ~ + ~ N
}
\end{equation}
is a fundamental neutrino interaction
which has never been observed.
The Standard Model cross section for
this process is given by:
\begin{eqnarray}
\label{eq::cohsm}
\rm{
( \frac{ d \sigma }{ dT } ) ^{coh} _{SM} }  
& = & 
\rm{
\frac{ G_F^2 }{ 4 \pi }
m_N  [ Z ( 1 - 4 \s2tw ) - N ]^2
[ 1 -  \frac{m_N T_N }{2 E_{\nu}^2 } ] 
} \nonumber \\
\rm{
\sigma _{tot} } 
 & = & 
\rm{
\frac{ G_F^2  E_{\nu}^2 }{ 4 \pi }
 [ Z ( 1 - 4 \s2tw ) - N ]^2 
}
\end{eqnarray}
where $\rm{m_N}$, N and Z are the mass, neutron number
and atomic number of the nuclei, respectively,
$\rm{E_{\nu}}$ is the incident neutrino energy
and $\rm{T_N}$ is the measure-able
recoil energy of the nucleus.
This formula is applicable at $\rm{E_{\nu} < 50~MeV}$
where the momentum transfer ($\rm{Q^2}$) is small such that
$\rm{ Q^2 R^2 < 1 }$, where R is the nuclear size.
Corrections due to nuclear form factors can be neglected
at low energies: $\rm{ Q^2 \ll m_N^2 }$.
The $\sim$N$^2$ enhancement on the cross-sections
signifies coherence.

\begin{figure}[hbt]
\begin{center}
\psfig{file=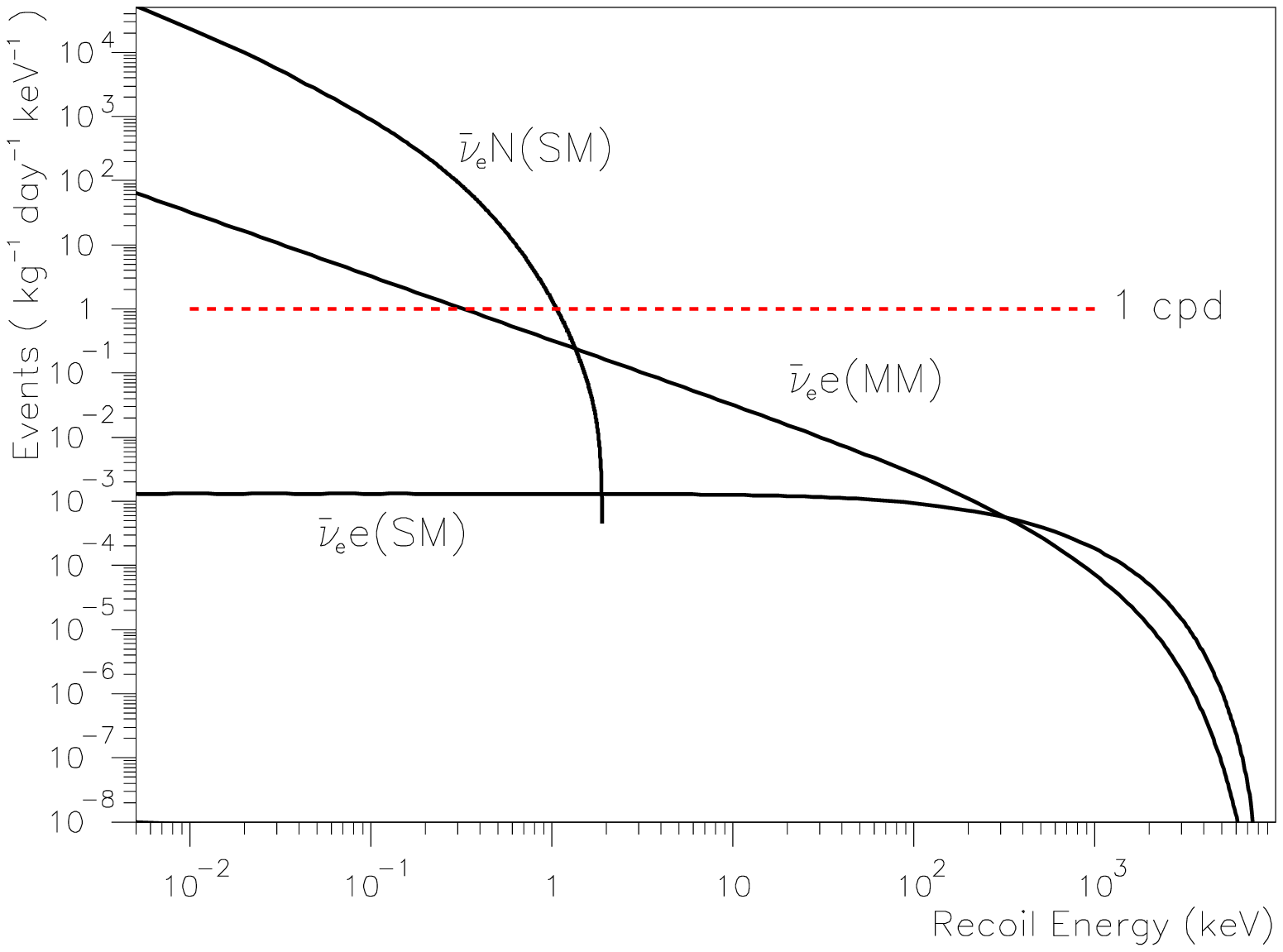,width=9.5cm}\\
\psfig{file=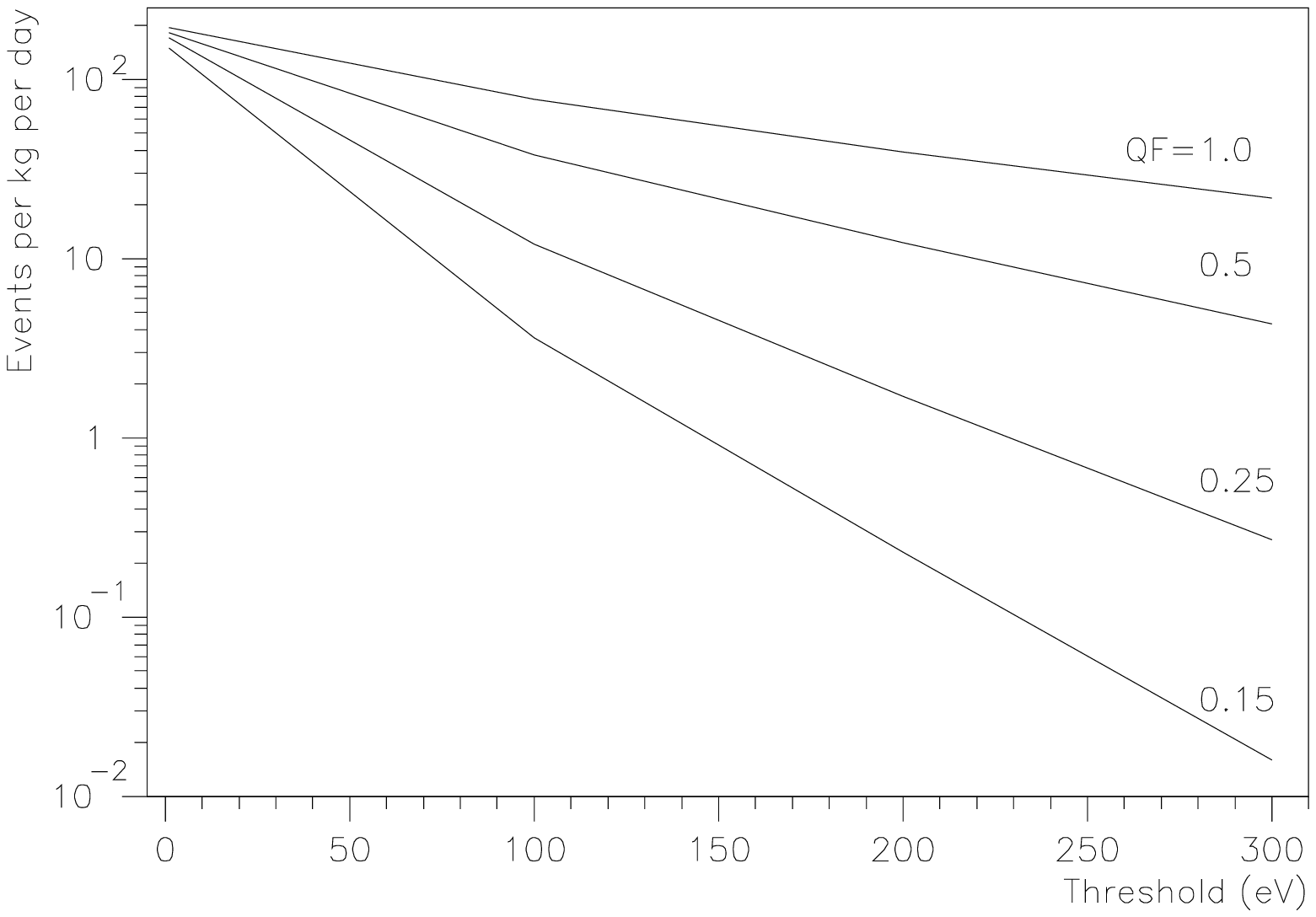,width=9.5cm}
\end{center}
\caption{
(a) Top: The differential cross-section of the various
neutrino interaction channels, at
KS-Lab with Ge as the target isotope.
The background level of 1~cpd is also
shown.
(b) Bottom: The variations of
the neutrino coherent scattering
event rates
versus threshold at different
quenching factors for
a 1~kg ULEGe detector at KS-Lab.
}
\label{diffcs}
\end{figure}

Measurement of the coherent scattering
cross-section would provide a sensitive
test to the Standard Model\cite{smtest,pibeam} probing the
weak nuclear charge and radiative corrections
due to possible non-standard neutrino interactions 
or additional neutral gauge bosons.
The coherent interaction plays important role
in astrophysical processes where the
neutrino-electron scatterings are suppressed
due to Fermi gas degeneracy. It is significant
to the neutrino dynamics and energy transport
in supernovae and neutrons stars\cite{astro}.
Being a new detection channel for neutrinos,
it may provide new approaches to study other aspects
of neutrino physics. For instance, the
detection of supernova neutrinos
using this mode was recently discussed\cite{nostos}.
Coherent scattering with the nuclei is also the
detection mechanism adopted in the
direct Dark Matter searches\cite{pdg06,cdmreview},
such that its observations
and measurements with the
known particle neutrino is an
important milestone as well.
Furthermore, neutrino coherent scattering
may be a promising avenue towards a compact
and relatively transportable neutrino detector, 
an application of which can be for
the real-time monitoring on the operation
of nuclear reactors\cite{monitor,2phase}, 
a subject of paramount global importance
in the non-proliferation of nuclear materials.



Nuclear power reactors are intense source of
electron anti-neutrinos ($\nuebar$) at the
MeV range, from which many important
neutrino experiments were based.
The $\nuebar$-spectra are
well-modeled, while good experimental
control is possible via the reactor ON/OFF comparisons.
The magnetic moment results at KS\cite{texonomagmom}
set the stage to pursue the studies of
neutrino-nucleus coherent scatterings
with reactor neutrinos\cite{texonocohsca}.

The maximum nuclear recoil energy
at momentum transfer much larger than neutrino masses
is given by:
\begin{equation}
\rm{
T_N^{max}  ~ = ~ \frac{2 E_{\nu}^2}{M_N + 2 E_{\nu}}
}
\end{equation}
The maximum neutrino energy
for the typical reactor $\nuebar$ spectra
is about 8~MeV, such that
$\rm{T_N^{max} = 1.9~keV  }$
for Ge target (A=72.6).
The differential cross section
for coherent scattering versus
nuclear recoil energy
with typical reactor $\nuebar$ spectra is
displayed in Figure~\ref{diffcs}a.
Overlaid for comparisons 
are the contributions due to 
neutrino-electron scatterings
from the Standard Model and magnetic moment
effects at the present limit,
as well as the 1~cpd benchmark background level.

In ionization detectors like HPGe,
the measure-able energy is only a
fraction of the energy deposit
for the nuclear recoil events
which have large dE/dx.
The {\it quenching factor (QF)},
defined as the ratio of the measure-able to
the deposit energy, is about 0.2-0.25 for Ge
in the $<$10~keV region\cite{geqf}.
Accordingly, the maximum measure-able energy
for nuclear recoil events in Ge due to reactor
$\nuebar$ is about 400-500~eV.  

The event rates
for neutrino-nucleus coherent scattering
at different threshold and quenching factors
at KS are depicted
in Figure~\ref{diffcs}b.
The goals of the R\&D program is to
devise an experiment  
based on Ge-detector with a mass 
range of 1~kg, 
a threshold as low as 100~eV, and
an on-site background level
of 1~cpd below 1~keV.
At the typical QF=0.25,
the event rate for such configurations
will be 11~kg$^{-1}$day$^{-1}$ or
4000~kg$^{-1}$yr$^{-1}$, at
a signal-to-background ratio of $>$22.

\subsection{Neutrino Magnetic Moments}

Neutrino magnetic moments ($\munu$) are parameters
characterizing the spin-dependent
couplings of the neutrinos to the photons\cite{munureview}.
A limit of 
\begin{equation}
\label{eq::limit}
\rm{
\munuebar  ~ <  ~ 7.4  \times 10^{-11} ~ \mu_B
}
\end{equation}
at 90\% confidence level (CL) has been achieved
at KS-Lab with a high-purity germanium detector at
a threshold of 12~keV and a background level
of 1~cpd\cite{texonomagmom}. 
This result was further improved upon 
by the GEMMA experiment\cite{gemma07} with a similar design 
but at a closer location to the reactor core.

A natural by-product of a detector with
kg-mass, 100~eV threshold and
1~cpd background level 
would be to further enhance
the sensitivities of
$\munu$-searches at reactors. 
The physics threshold can be as low as 
500~eV just
above the coherence scattering detectable
energy cut-off.
A finite $\munu$
will contribute to $\nu$-e scattering
with a differential cross-section term
given by:
\begin{equation}
\label{eq::mm}
\rm{
( \frac{ d \sigma }{ dT } ) _{\munu}  ~ = ~
\frac{ \pi \alpha _{em} ^2 {\it \munu } ^2 }{ m_e^2 }
 [ \frac{ 1 - T/E_{\nu} }{T} ] ~ .
}
\end{equation}
The effects are much
enhanced at this sub-keV energy due to 
the 1/T dependence,
as illustrated in Figure~\ref{diffcs}a.
An improved sensitivity range down
to $\rm{\sim 10^{-11} ~ \mu_B}$ 
can be expected.

\subsection{Cold Dark Matter Searches}

There are compelling evidence that about 25\% of the 
energy density in the universe is
composed of Cold Dark Matter\cite{pdg06,cdmreview} due to
a not-yet-identified particle, generically categorized
as Weakly Interacting Massive Particle 
(WIMP, denoted by $\chi$).
A direct experimental detection of WIMP 
is one of the biggest challenges in frontier
astroparticle physics. 

The WIMP will interact
with matter pre-dominantly
via the same coherent scattering
mechanism like the neutrinos:
\begin{equation} 
\rm{
\chi ~ + ~ N ~ \rightarrow ~
\chi ~ + ~ N ~ .
}
\end{equation}
The major difference is that cosmology dictates
that the WIMPs should be massive and its motion
non-relativistic to
be consistent with the observational data on
structure formation\cite{pdg06}.
In addition, it is possible to have 
a spin-dependent
interaction between WIMP and matter.

Supersymmetric (SUSY) particles\cite{pdg06}
are the leading WIMP candidates.
The popular SUSY models prefer 
WIMP mass ($\mwimp$) of the range
of $\sim$100~GeV,
though light neutralinos
remain a possibility\cite{lightsusy}.
Most experimental programs
optimize their design in the high-mass region
and exhibit diminishing sensitivities for $\rm{\mwimp < 10 ~GeV}$,
where there is an allowed region 
if the annual modulation data 
of the DAMA experiment\cite{damaallowed}
are interpreted as WIMP signatures.
Simple extensions of the Standard Model with a 
singlet scalar favors light WIMPs\cite{smscalar}.
To probe the low-mass region, detector with sub-keV
threshold is necessary. 
Such threshold will also allow the studies of
WIMPs bound in the solar system\cite{solarwimp},
and non-pointlike SUSY candidates like Q-balls\cite{qball},
but presents a formidable
challenge to detector technology and to background control.

\begin{figure}
\begin{center}
\psfig{file=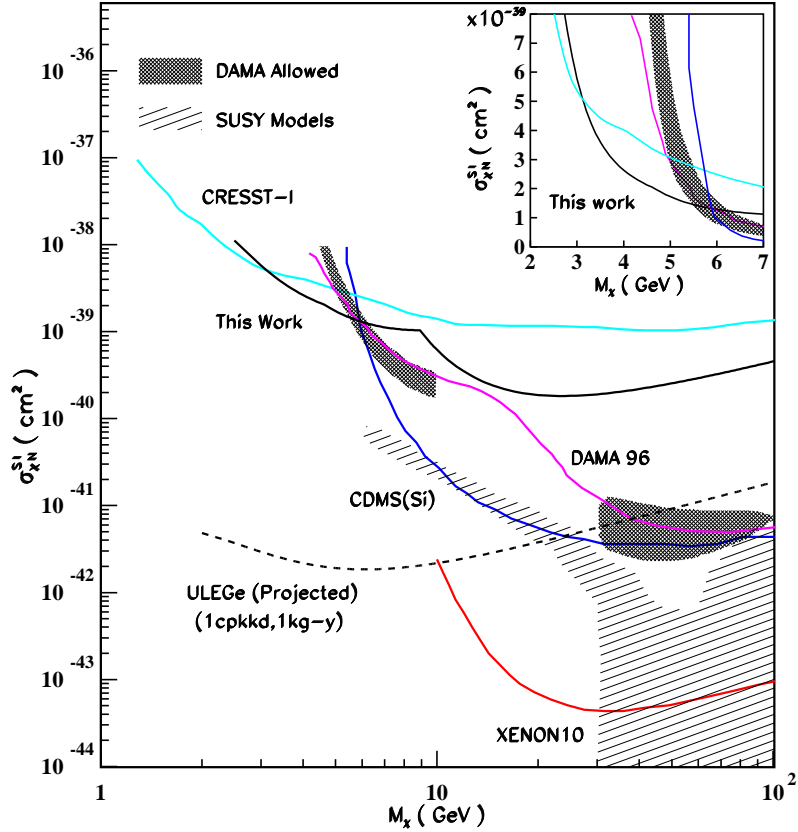,width=10.5cm}
\caption{
\label{cdmplot}
Exclusion plots of spin-independent
$\chi$N cross-section versus
WIMP-mass, displaying the
KS-ULEGe limits and
those from various experiments
defining the current boundaries.
The DAMA allowed regions
are superimposed.
The striped region is that favored
by SUSY models.
Projected sensitivities of full-scale
experiments are indicated as dotted lines.
The relevant region is presented with
linear scales in the inset.
}
\end{center}
\end{figure}

In direct WIMP search experiments, the observed
nuclear recoil rate 
per unit detector mass ($\rm{ dR/dT_N }$) 
at recoil energy $\rm{T_N}$ 
is related to 
the elastic WIMP-nucleon cross section ($\sigma _{\chi N}$)
and $m_{\chi}$ via\cite{cdmmaths}:

\begin{equation}
\label{eq::cswimp}
{\rm 
\frac{dR}{dT_N}   ~  = ~  
[   \frac{\rho_{\chi}}{ 2 ~  m_{\chi}\mu^2_{N}}  ] ~ 
[ ( \frac{\mu _A ^2 }{ \mu _N ^2 } ) ( \frac{C_A}{C_N} ) 
F^2 ( q )  \sigma _{\chi N} ] } ~
          \int_{v_{min}}  \frac{\fut}{u}  ~  d \, ^3 u  ~ ,
\end{equation}
where
$\rm{\mu _{A/N}}$ denotes 
the reduced mass of the target
nucleus/nucleon with respect to the WIMPs,
while
$\rm{ F ( q )}$ is the nuclear form factor
where $\rm{ q =\sqrt{(2 M_A T_N)}}$ is the
nucleus recoil momentum,
$\rho_{\chi} \sim 0.3 ~ {\rm GeV cm^{-3}}$ is
the standard local (solar neighborhood) halo WIMP density, 
$\fut$ is the time-dependent
distribution of the WIMP velocity $\vec{u}$ relative to the 
detector typically assumed to be
Maxwellian with a cut-off at the galactic escape velocity. 
The lower bound of the integral
$
 v_{min} =
{\rm
\sqrt{ ( M_A T_N ) / ( 2 \, \mu^2_{\chi} ) }  
}
$
denotes the minimal WIMP velocity necessary to 
kinematically induce a nuclear recoil interaction.
The enhancement factor from nucleon to nucleus 
is denoted by $\rm{ ( {C_A}/{C_N} ) }$, 
which is equal to $\rm{A^2}$ due to coherent effects
for spin-independent couplings.
In the spin-dependent case\cite{tovey}, couplings of the WIMPs
to protons 
and neutrons 
are in general different, depending on 
\begin{equation}
\rm{
C_A = ( \frac{8}{\pi}  ) \, 
[ \,  a_{\, p}  < S_p > + a_{\, n} <S_n >  \, ]^2 
( \frac{J+1}{J} )  ~~ ; ~~ 
C_{p/n} = ( \frac{6}{\pi} ) \, a_{\, p/n}^2
}
\end{equation}
where 
J is total nuclear spin,
$\rm{a_{\, p/n}}$ are the WIMP-proton/neutron coupling constants
and $\rm{<S_{p/n}>}$ are the expectation values of 
the proton/neutron spins within the nucleus\cite{spinvalue}. 
At the typical range of $m_{\chi} \sim 10 ~ {\rm GeV}$,
the maximal WIMP velocity of $ v \sim 0.002 {\rm c}$
corresponds to a kinetic energy of $\rm{T_{\chi} \sim 20 ~ keV}$
and the maximal recoil energy at Ge of
$\rm{T_N \sim 5 ~ keV}$.

The uniqueness and advantages of having a low-threshold detector
are twofold. 
Firstly, WIMPs with lower masses become detectable at a
lower threshold, as indicated in Eq.~\ref{eq::cswimp},
thereby opening a new window of observation.
Secondly, the lower bound of the integral $\rm{v_{min}}$ 
in Eq.~\ref{eq::cswimp},
which represents the
minimum velocity of a WIMP that produces a recoil energy 
$\rm{T_N}$, is proportional to $\rm{T_N}$. 
Therefore, a lower threshold
allows a larger range of WIMPs to contribute in an observable 
interaction and
hence results in better sensitivities for all values of $m_{\chi}$.
 
A summary of the sensitivities 
of the spin-independent cross-section 
($\csnospin$) versus $m_{\chi}$ 
is depicted in Figure~\ref{cdmplot}.
The best sensitivities to date in the 
low WIMP mass region ($m_\chi < 10 ~{\rm GeV}$)
are due to data taken at KS
with a ultra-low-energy germanium 
prototype\cite{texonowimp} with threshold
of 200~eV, superseding those
from the previous CRESST-I experiment\cite{cresst1}
with sapphire($\rm{Al_2 O_3}$)-based cryogenic detector
at a threshold of 600~eV.
The other experiments
defining the current boundaries\cite{cdmbounds}
as well as the DAMA-allowed regions\cite{damaallowed}
are superimposed.
The striped region is that favored
by SUSY models\cite{lightsusy}.
Projected sensitivities of full-scale
experiments are indicated as dotted lines.
 
\section{Ultra-Low-Energy Germanium Detector}
\label{sect::ulege}

Although the $\nu$N and $\chi$N
cross-section of Eq.~\ref{eq::cohsm}
is relatively large
due to the respective
N$^2$ and A$^2$ enhancement by coherence,
the small kinetic energy from nuclear recoils
poses severe experimental challenges
both to the detector sensitivity and to background
control. 
Various detection schemes
using cryogenic\cite{cryo}, gas\cite{nostos,mpgd} 
or
liquid-gas two phase\cite{2phase} detectors
have been investigated taking
reactor neutrinos as the source. 
There was a recent study on using
neutrino beam from 
stopped-pion facilities\cite{pibeam}.
Alternatively,
germanium-based ionization detectors\cite{texonocohsca,chicago}
offer a more matured technology,
less costly to build and to operate,
more compact, and easier to scale-up.
This detector technology  
has been widely and successfully
used in various areas of low energy neutrino physics
and cold dark matter searches.

``Ultra-Low-Energy'' Germanium (ULEGe)
detectors, developed originally for soft X-rays
detection, are candidate technologies to meet
these challenges of probing into the previously
unexplored low-energy domain.
These detectors typically have
modular mass of 5-10~grams while detector array
of up to N=30 elements have been successfully
built. 
Various ULEGe prototypes\cite{ulege} were constructed
in the course of our R\&D program,
the highlights of which are discussed in the
following sections.
Complementary to these, measurements of
QF in Ge at the sub-keV
range is pursued at a neutron
facility, following the previous
successful measurements of QF in
CsI(Tl)\cite{qfcsi}. 

\begin{figure}[hbt]
\begin{center}
\psfig{file=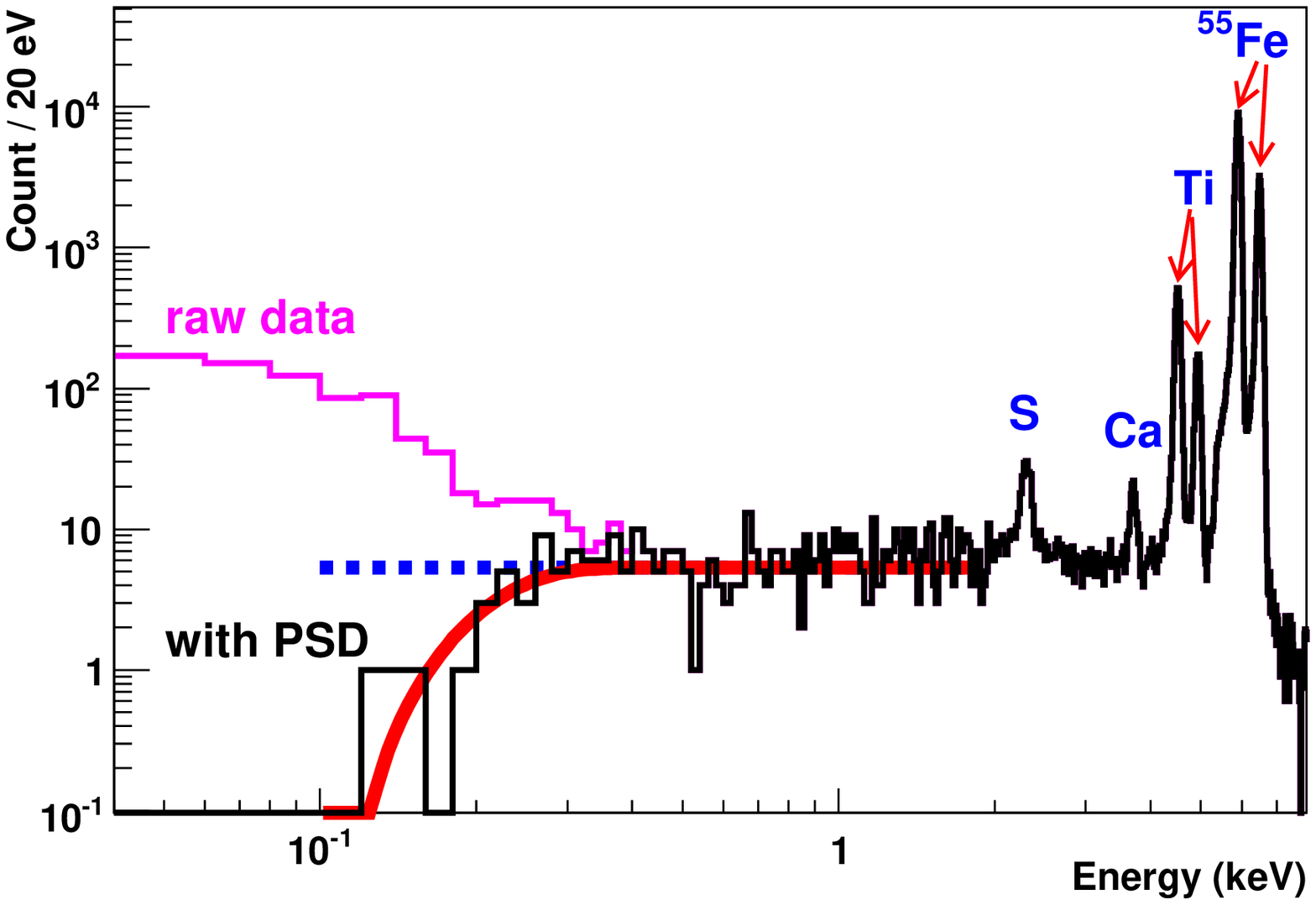,width=10cm}\\[5ex]
\psfig{file=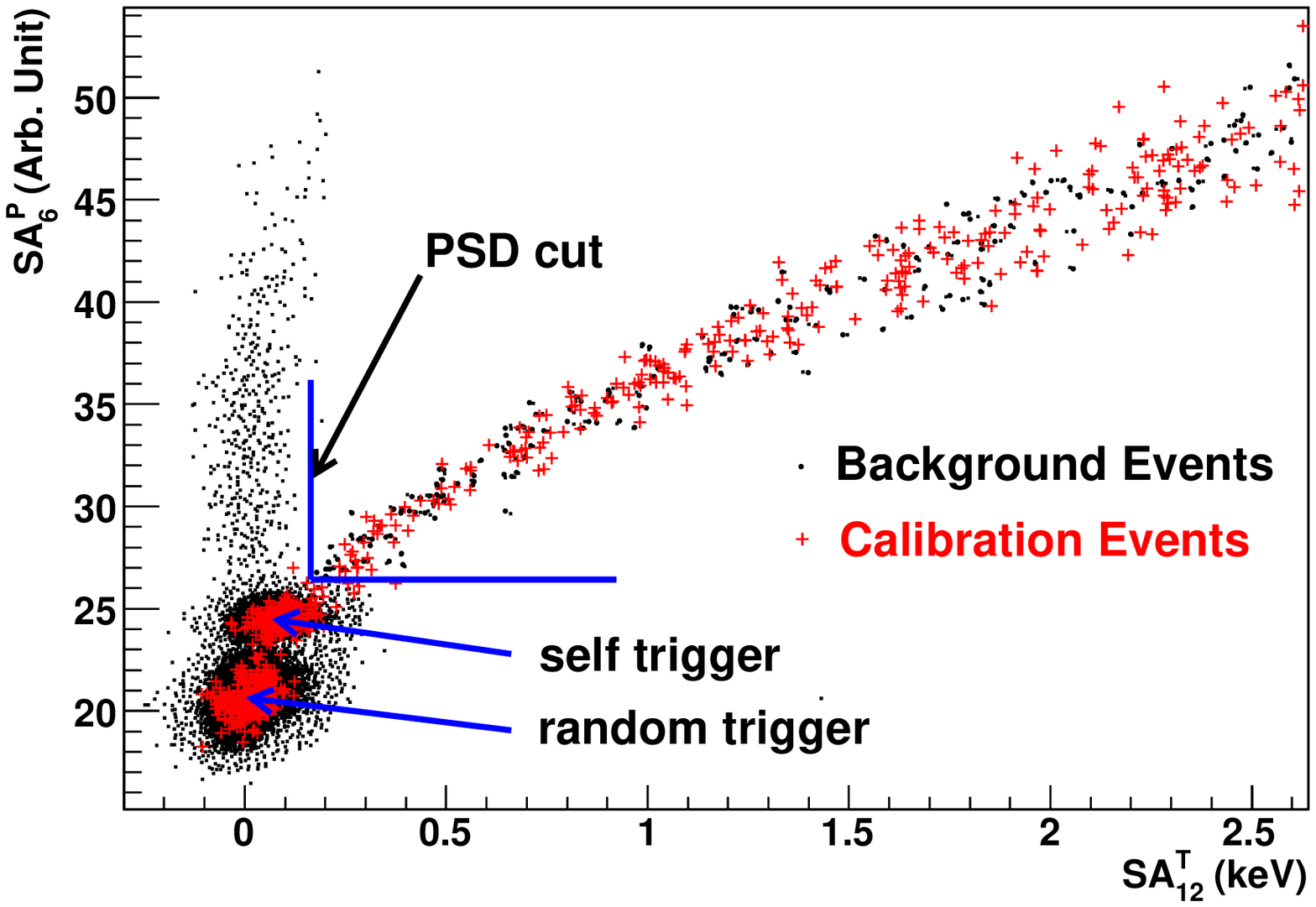,width=10cm}
\caption{
\label{fe55}
(a) Top: Measured energy spectrum of the ULEGe
with $^{55}$Fe
source showing also X-ray peaks
from various materials.
The best-fit of the spectrum
at $\rm{0.5 - 2 ~ keV}$ is extrapolated
to low energy.
The black histogram represents events
selected by PSD cuts.
Deviations from the expected spectra
contribute to PSD efficiencies.
(b) Bottom: Scattered plots of the two signals with
different shaping times and amplification factors,
which are
constrained by both energy and timing,
for both calibration and physics events.
The PSD selection is shown.
}
\end{center}
\end{figure}

\subsection{Detector Threshold}

A typical measured energy spectrum
ULEGe prototype is depicted 
in Figure~\ref{fe55}a.  
Pulsed optical feedback preamplifiers
were used to extract the signals
from the electron-hole pairs.
The output was transferred to 
two amplifiers at
6~$\mu$s and 12~$\mu$s shaping time,
and read out by a 20~MHz Flash Analogue
Digital Convertor\cite{eledaq}.
The sampling period extended from
t=-20~$\mu$ to 70~$\mu$s 
where t=0 was defined by the trigger instant.
Calibration was achieved by
external $^{55}$Fe X-ray sources (5.90 and  6.49~keV)
together with X-rays from
titanium (4.51 and 4.93~keV), calcium (3.69~keV)
and sulphur (2.31~keV). 
Excellent linearity down to threshold 
was demonstrated with a test-pulser.
A random trigger uncorrelated to the detector
provided the dead time and efficiencies measurements
as well as the zero-energy calibration.
The electronic noise which defines the 
threshold can be suppressed by
pulse shape discrimination (PSD)
through correlating the two output
signals at the two shaping times, 
as illustrated in Figure~\ref{fe55}b. 
A threshold of 200~eV 
at $\sim$50\% selection efficiency\cite{texonowimp}
was achieved.

\subsection{Background Measurement}

\begin{figure}[hbt]
\begin{center}
\psfig{file=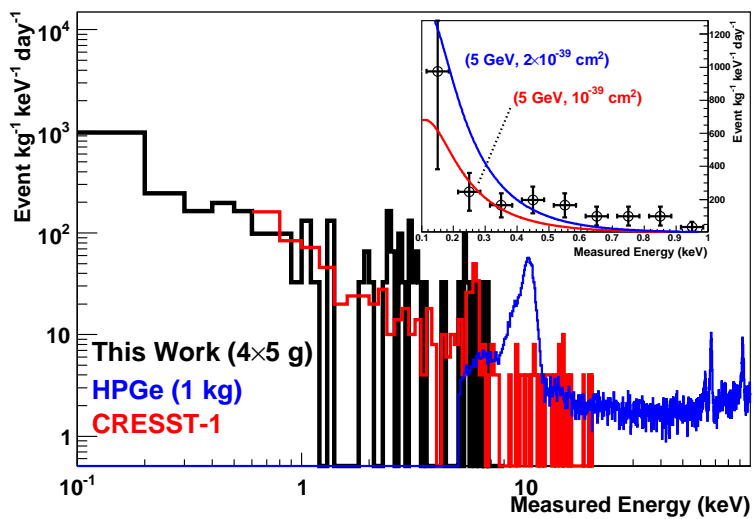,width=10cm}
\caption{
The measured spectrum of ULEGe with
0.338~kg-day of data,
after cosmic-ray and anti-Compton vetos
as well as PSD selections.
Background spectra of the CRESST-I experiment
and the HPGe
are overlaid for comparison.
The expected nuclear recoil spectra
for two cases
of $\rm{ ( \mwimp , \csnospin ) }$
are superimposed onto the spectrum
shown in linear scales in the inset.
}
\label{ulegebkg}
\end{center}
\end{figure}

Background studies were made with
the prototypes 
at the KS reactor laboratory and
at the Yangyang Underground laboratory (Y2L)
in South Korea. 
The detectors were surrounded by active
anti-Compton crystal scintillator detectors,
passive shieldings as well as active cosmic-ray
veto scintillators $-$
identical to the those for the KS
1~kg HPGe experiment\cite{texonomagmom}.
Data were taken under different hardware and
software configurations.
The best spectrum in terms of background and
threshold is shown 
in Figure~\ref{ulegebkg},
with those from CRESST-1\cite{cresst1}
and the 1-kg HPGe\cite{texonomagmom} 
superimposed.
The background was translated to 
the WIMP exclusion regions\cite{texonowimp} 
in Figure~\ref{cdmplot} via Eq.~\ref{eq::cswimp}.
The KS ULEGe data define the exclusion
boundary for $\rm{\mwimp \sim 3 - 6 ~ GeV}$.
Further work on background understanding
and suppression are under way.

\subsection{Detector Mass Scale-Up}

\begin{figure}[hbt]
\begin{center}
\psfig{file=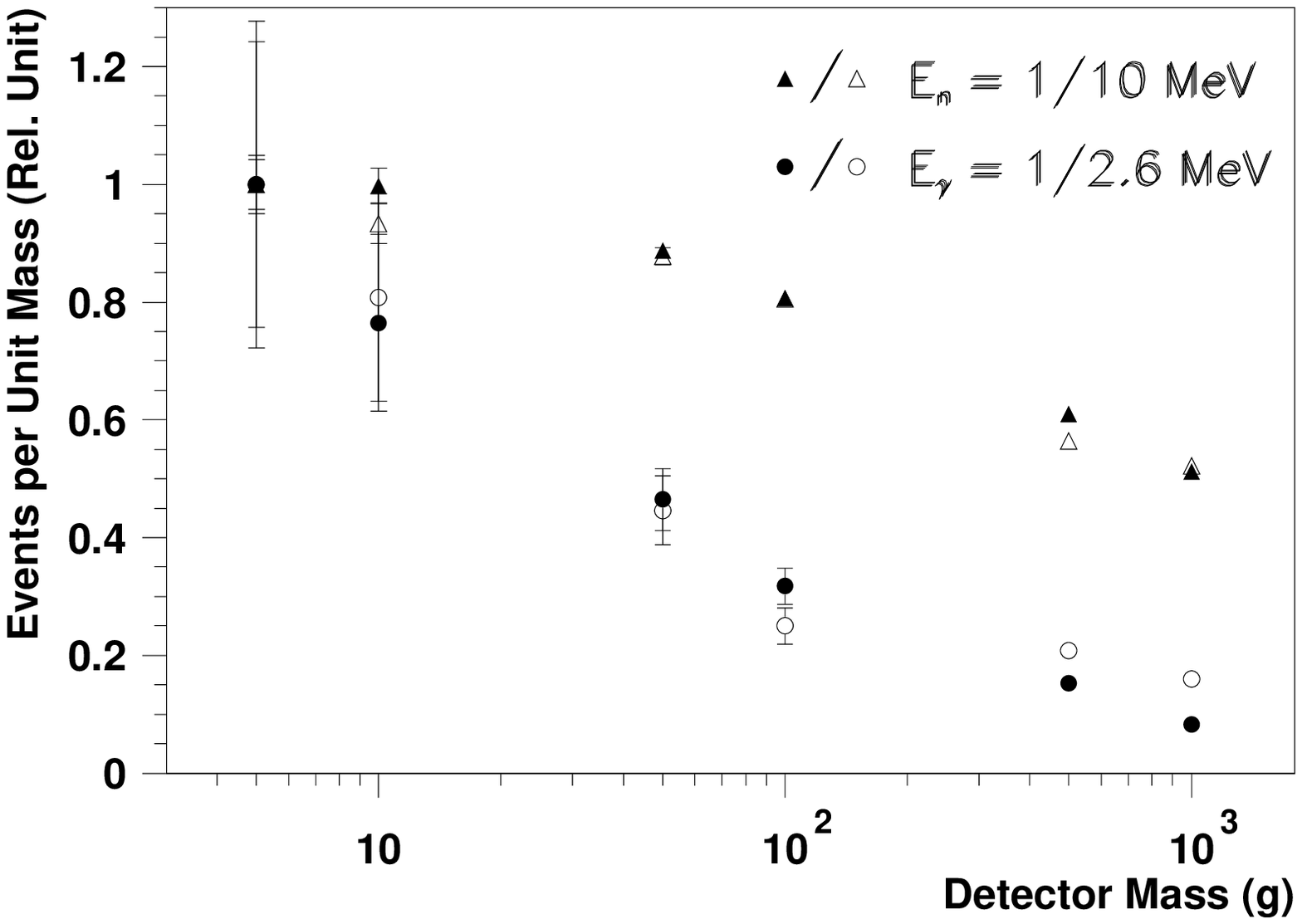,width=9.5cm}\\
\psfig{file=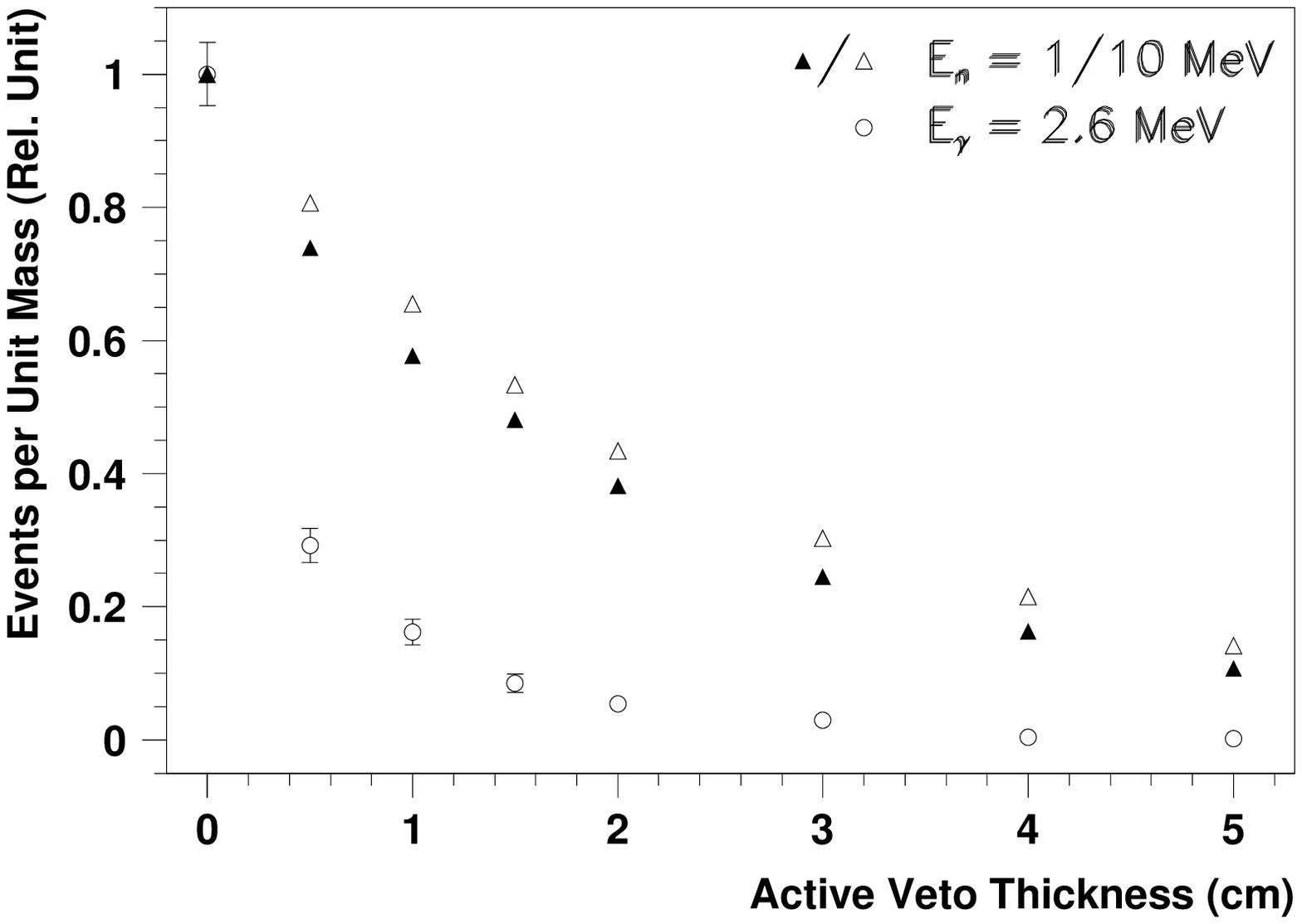,width=9.5cm}
\end{center}
\caption{
\label{bkgmc}
(a) Top: Simulated background rates
at the sub-keV range
per unit mass versus detector
mass at the same external $\gamma$-background
level.
(b) Bottom: Background reduction factors
in a 1-kg detector
versus thickness of 4$\pi$ Ge-veto layer.
Photon energy of 1.0 and 2.6~MeV 
and neutron energy of 1.0 and 10~MeV were  adopted.
}
\end{figure}

The ULEGe element can be constructed
in multi-array configuration.
Alternatively, ``segmented'' ULEGe detector
read out as pixels in integrated sensors
have been constructed and is being studied.
Moreover, inspired by a novel 
design in the 1980's\cite{luke},
there are recent important advances in
the construction of 
a single-element ULEGe of 500-g mass\cite{chicago},
with which a detector threshold of 300~eV was demonstrated.
This offers great potentials of scaling-up the detector
mass to the kg-range and beyond with simple 
electronics and data acquisition schemes.

Realistic simulations were performed on
the behaviour of photons and neutrons
with kg-scale ULEGe. 
The variations of the ``after-cut'' rates
the with detector mass 
are depicted in Figure~\ref{bkgmc}a.
The mass-normalized count rates at low energy ($<$10~keV)
under a constant external $\gamma$-background
decrease by a factor of 10
for increasing Ge target mass 
from 5~g to 1~kg,
due to self-absorption of the target. 
Similar behaviour applies to neutron-induced
background, though the difference is less.
Illustrated in Figure~\ref{bkgmc}b are the 
background reduction factors 
for a 1~kg detector
versus the thickness of 
possible Ge-veto layers. 
The potential reach on $\csnospin$ sensitivities
with a 1-kg detector
at 1~cpd background is depicted in 
Figure~\ref{cdmplot}.

\section{Outlook}

A detector with 1~kg mass, 100~eV threshold
and 1~cpd background level has important
applications in neutrino and dark matter physics,
as well as in the monitoring of reactor operation. 
Crucial advances have been made in 
adapting the ULEGe technology to meet these
challenges. Competitive limits have been
achieved in prototype studies on the
WIMP dark matter spin independent couplings.
Intensive research programs are being
pursued along these fronts towards
realization of such experiments.

\section*{Acknowledgments}

The author is grateful to the CosPA-2007
Symposium for the opportunity to present
this report.
This work is supported by contracts 
NSC94-2112-M-001-028 
and 
NSC 96-2112-M-001-005
from the National
Science Council, Taiwan, as well as
by the Pilot Project Scheme 2004-06
from the Academia Sinica, Taiwan.
The Y2L-related work is supported
by funding support 
from the National Natural Science Foundation,
China through contract 10620140100.
The QF measurement is supported by
neutron facility at the China Institute
of Atomic Energy. The infrastructure
support at Y2L is provided by KIMS Collaboration,
South Korea.

\end{document}